\documentclass[a4paper,11pt,amsmath,amssymb]{article}
\pdfoutput=1 

\usepackage{jheppub} 

\usepackage[T1]{fontenc} 

  \usepackage{graphicx}
\usepackage{amsfonts}
\usepackage{color} 
\usepackage[colorlinks=true,urlcolor=blue,linkcolor=blue,citecolor=blue]{hyperref}
\usepackage[T1]{fontenc}
\usepackage{lmodern} 

\title{\boldmath Reanalysis of lattice QCD spectra leading to the $D_{s0}^*(2317)$ and $D_{s1}^*(2460)$ }
\date{\today} 
\author{A. Mart\'inez Torres,}
\affiliation{Instituto de F\'isica, Universidade de S\~ao Paulo, C.P. 66318, 05389-970 S\~ao 
Paulo, SP, Brazil.}
\author{E. Oset,}
\affiliation{Departamento de
F\'{\i}sica Te\'orica and IFIC, Centro Mixto Universidad de
Valencia-CSIC Institutos de Investigaci\'on de Paterna, Aptdo.
22085, 46071 Valencia, Spain.}
\author{S. Prelovsek}
\affiliation{Department of Physics at University of Ljubljana, Jozef Stefan Institute, 1000 Ljubljana, Slovenia.}
\author{A. Ramos}
\affiliation{Departament d'Estructura i Constituents de la Mat\`eria and Institut de Ci\`
encies del Cosmos, Universitat de Barcelona, Mart\'i i Franqu\`es 1, E-08028 Barcelona, Spain}
 
\emailAdd{amartine@if.usp.br}
\emailAdd{Eulogio.Oset@ific.uv.es}
\emailAdd{sasa.prelovsek@ijs.si}
\emailAdd{ramos@ecm.ub.edu} 
 
\abstract{
We perform a reanalysis of the energy levels obtained in a recent lattice QCD simulation, from where the existence of bound states of $KD$ and $KD^*$ are induced and identified with the narrow $D_{s0}^*(2317)$ and $D_{s1}^*(2460)$ resonances. The reanalysis is done in terms of an auxiliary potential, employing a single-channel basis {$KD^{(*)}$}, and a two-channel basis {$KD^{(*)}, \eta D_s^{(*)}$}. By means of an extended L\"uscher method we determine poles of the continuum $t$-matrix, bound by about 40 MeV with respect to the $KD$ and $KD^*$ thresholds, which we identify with the $D_{s0}^*(2317)$ and $D_{s1}^*(2460)$ resonances. Using a sum rule that  reformulates  Weinberg  compositeness condition we can determine that the state $D_{s0}^*(2317)$ contains a $KD$ component in an amount of about 70\%, while the state $D_{s1}^*(2460)$ contains a similar amount of $KD^*$. We argue that the present lattice simulation results do not still allow us to determine which are the missing channels in the {bound state} wave functions and we discuss the necessary information that can lead to answer this question. 
}

\begin{document} 
\maketitle
\flushbottom

\section{Introduction}
{The scalar $D_{s0}^*(2317)$ and axial $D_{s1}(2460)$  mesons were experimentally found slightly below $K D$ and $K D^*$ thresholds \cite{Aubert:2003fg,Besson:2003cp,Lees:2011um,pdg}.    These are one of the few shallow  bound states in the meson sector, and therefore deserve special attention. The effect of  thresholds was recently considered using lattice QCD for the first time in this system in \cite{sasa,sasa1}, where interpolators of $KD$ and $KD^*$ type have been employed in addition to $\bar sc$ ones.  The $N_f\!=\!2+1$ simulation obtained three energy levels for $m_\pi\simeq 156~$MeV in the $KD$ and $KD^*$ systems .}  
The fact that these levels appear clean with the $KD$ and $KD^*$ interpolators, together with the observation that the lowest one appears below and not far from the corresponding $KD$ or $KD^*$ threshold, hint to a possible molecular structure for this state.
The scattering length and the effective range were determined from the two lowest energy levels  in \cite{sasa,sasa1} and, using the effective range formula, bound states were  found at about 40 MeV below the  respective $KD$ and $KD^*$ thresholds. These were identified with the scalar $D_{s0}^*(2317)$ and axial $D_{s1}^*(2460)$ states respectively.  

Actually, the two lower levels employed in the analysis of  \cite{sasa,sasa1}
are separated by 130 MeV, which makes the use of the effective range formula a bit extreme,  and the information of the upper level was disregarded. In the present work we perform a reanalysis of these lattice spectra which does not rely upon the effective range formula and takes advantage of the information of the three levels. The analysis is done using the auxiliary potential method \cite{misha}, equivalent to 
the one of L\"uscher \cite{luscher} in single or coupled channels, but allowing also to obtain phase-shifts for arbitrary energies. The lattice simulations are particularly suited for this kind of study because, for the same value of $L$, they  produce several energy levels which provide information on the energy dependence of the potential needed to interpret the spectra. 

We first perform a single channel analysis, with $KD$ or $KD^*$, which permits determining the two parameters of an energy dependent potential from a fit to the three energy levels of the box.  This potential is then used in the continuum, leading to poles of the $KD$ and $KD^*$ scattering amplitudes, which lie about 40 MeV below the respective thresholds. A reformulation of the Weinberg compositeness condition \cite{composite,baru} is then used to determine the amount of $KD$ and $KD^*$  in the respective wave functions.  A different method to learn about the amount of meson component, or equivalently the amount of non-meson component, $Z$, in the wave function, is from the dependence of the spectrum on the twisting angle, imposing twisted boundary conditions on the fermion fields \cite{Agadjanov:2014ana}.

 The compositeness condition was extended leading to a new sum rule in an arbitrary number  of coupled channels \cite{danijuan}, which is reformulated in \cite{hyodojido,sekiward,hyodohosaka,hyodoreport,sekirep} for the case of energy dependent potentials. The sum rule contains two terms (see Eq. (133) in \cite{hyodoreport}), one involving the derivative of the two-particle loop function, which is identified with the probability of the state containing this particular two-particle component of the coupled channels. {The second term involves} the derivative of the potential with respect to the energy, which accounts for the probability of the state to be in other components not explicitly considered in the approach{, for example omitted two-meson channels or $\bar qq$}. An illustrative example is given in \cite{acetidelta}, where one starts from a two channel problem with energy independent potentials which generate dynamically a bound state. The problem is then reformulated in terms of one channel and an effective potential, which however becomes energy dependent. This allows one to see that the term in the sum rule involving the derivative of the loop function accounts for the probability of the channel retained, while the term involving the derivative of the potential accounts for the probability of the omitted channel.

Having this in mind, we repeat the analysis of the lattice results using a two channel basis, involving  $KD, \eta D_s$ for $D_{s0}^*(2317)$ and $KD^*, \eta D_s^*$   for $D_{s1}^*(2460)$. The choice of channels relies on the results of coupled channels unitary approaches \cite{Kolomeitsev:2003ac,Guo:2006fu,dani,daniaxial,guohan,wang,hanhart_orginos,Cleven:2014oka,geng_weise,alten_geng}, 
which found those channels to be the relevant ones (in what follows we will mainly refer to Refs. \cite{dani,daniaxial} when we give details of the coupled-channel formulation).  Alternative scenarios for a non ${\bar q}q$ structure of these states have been also given \cite{van,Barnes:2003dj,nowak_rho,Nielsen:2009uh,Brambilla:2010cs,Esposito:2014rxa}.
With two channels and three energy levels one is forced to treat the three components of the coupled-channel potential  ($V_{11}, ~V_{12}, ~V_{22}$) as being energy independent. We observe that a fit to the energy levels is not possible in this case, indicating that these levels carry no information on the $\eta D_s$ and $\eta D_s^*$ channels. This can be explained since no interpolators of this type were used in \cite{sasa}, while it was also found there that the levels obtained were tied to the interpolators used. Further lattice information will be needed in the future to make progress in this direction and learn more about the components that build up the $D_{s0}^*(2317)$ and $D_{s1}^*(2460)$ wave functions. 

    With the available limited lattice information, we can {confirm that the} bound states of $KD$ and  $KD^*$  can be associated to the $D_{s0}^*(2317)$ and $D_{s1}^*(2460)$ states.   {We also confirm that these bound states   are mostly of $KD$ or  $KD^*$ nature,   estimating about 70 \% the probability of these components in their respective wave function. }    
 
The compositeness of the $D_{s0}^*(2317)$ based on indirect lattice data was first discussed in \cite{hanhart_orginos}, but employing a different method.  The scattering lengths of other scattering channels, free from disconnected diagrams, were obtained on the lattice and used to determine the parameters of their effective field theory, which was subsequently used to indirectly determine the scattering  parameters of $KD$ scattering and the pole position in this channel. Similarly, the scattering lengths from the simulation of Ref.~\cite{hanhart_orginos} were employed in \cite{geng_weise,alten_geng} to fix the low-energy constants of a covariant chiral unitary theory, which was then used to also identify, as composite states, the heavy-quark spin and flavour symmetry counterparts of  the $D^*_{s0}$.
    
As mentioned above, additional lattice information could help us improve our knowledge on the additional building blocks that these states might have. Indeed, preliminary spectra for these channels obtained including $KD$, $\bar sc$  as well as $\eta D_s$ interpolating fields have been presented in \cite{Ryan:lat14}.  Their plan is to perform a two-coupled channel analysis using a parametrization of the scattering matrix on the energy. This strategy has recently lead to the first results of the two-coupled channel  system $K\pi-K\eta$ from lattice QCD; the pole positions of the scattering matrix were subsequently found and related to the strange mesons \cite{Dudek:2014qha}. The approach presented here offers an alternative way to extract physical information from the lattice spectra in the future.

\section{Compositeness of states}
\label{section2}
We collect here the essential expressions relevant to  interpret the nature of hadrons generated dynamically from a given meson-meson interaction. 
Let us take two mesons ($K$ and $D$ for example) and an interacting potential $V$. The Lippmann-Schwinger equation produces the scattering amplitude $T$
\begin{align}
T=V+VGT,\label{BS}
\end{align}
where $G$ stands for the two meson propagator. We shall take relativistic propagators and Eq.~(\ref{BS}) will be the Bethe-Salpeter equation. The on-shell factorization of $V$ and $T$ allows one to convert Eq.~(\ref{BS}) into an algebraic equation with $G$ given by
\begin{align}
G=i\int\frac{d^4q}{(2\pi)^4}\frac{1}{q^2-m^2_1+i\epsilon}\frac{1}{(P-q)^2-m^2_2+i\epsilon},
\end{align}
where $P$ is the total two meson momentum. This factorization was justified in Refs.~\cite{nsd,ollerulf} by using dispersion relations in which the smooth energy dependent contribution of the left-hand-side cut was replaced by a constant in the region of interest.
The energy dependence was shown to be particularly weak in the case of the meson-baryon interaction \cite{ollerulf} due to the large baryon mass and, consequently, it will be even weaker in the present case due to the larger mass of the $D$ and $D^*$ mesons. 
The neglect of the left hand cut is also inherent in the L\"uscher formalism, as we shall see in Sect.~\ref{aux}. 

 Upon integration of the $q^0$ variable the loop function becomes
\begin{align}
G=\int\frac{d^3q}{(2\pi)^3}I(\vec{q}\,),\quad\quad I(\vec{q}\,)&=\frac{\omega_1(\vec{q}\,)+\omega_2(\vec{q}\,)}{2\omega_1(\vec{q}\,)\omega_2(\vec{q}\,)\left[P^2-(\omega_1(\vec{q}\,)+
\omega_2(\vec{q}\,))^2+i\epsilon\right]} \ ,
\label{eq:loop}
\end{align}
where $\omega_i(\vec{q}\,)$ is the meson on-shell energy. The loop function
must be conveniently regularized with a cut-off $q_\textrm{max}$, or employing dimensional regularization techniques.

Assume now the Bethe-Salpeter equation projected over $S$-wave and $V$ an energy independent potential in one channel (say $KD$). We then have
\begin{align}
T(1-VG)=V,\quad T=\frac{V}{1-VG}=\frac{1}{V^{-1}-G}.\label{T}
\end{align}
Let us now assume that the interaction $V$ produces a bound state, which we will refer to as a two meson composite state or a dynamically generated state. {We shall see that the energy independent potential can not lead to a genuine state, for example a $\bar qq$ state with a weak coupling to two mesons}. In the case of one channel, the coupling $g$ of the bound state is obtained by requiring that around the pole $s=s_0$ (with $s=P^2$ being a Mandelstam variable)
\begin{align}
\label{gsq}
T\sim\frac{g^2}{s-s_0},\quad{\rm hence:}~ ~ g^2=\lim_{s\to s_0} (s-s_0)T.
\end{align}
Since $V^{-1}-G=0$ at the bound state pole, we find in the case of an energy independent potential using L'Hopital's rule
\begin{align}
g^2=\frac{1}{-\frac{\partial G}{\partial s}},\quad -g^2\frac{\partial G}{\partial s}=1\label{LH}.
\end{align}

The property of Eq.~(\ref{LH}) can be generalized to coupled channels and, in the case of an energy independent potential (and two channels), one finds:
\begin{align}
V&=\left(\begin{array}{cc}V_{11} &V_{12}\\V_{12}&V_{22}\end{array}\right),\quad G=\left(\begin{array}{cc}G_1&0\\0&G_2\end{array}\right),\label{V2ch}\\
T&=(1-VG)^{-1}V, \label{T2ch}\\
\quad g_i g_j&=\lim_{s\to s_0}(s-s_0)T_{ij},\quad \sum_i\left(-g^2_i\frac{\partial G_i}{\partial s}\right)=1\label{g}
\end{align}
Equation~(\ref{LH}) is a reformulation of the Weinberg compositeness condition~\cite{composite}, which is usually applied to loosely bound states, meant to be used at higher binding energies, while Eq.~(\ref{g}) is the extension to many coupled channels~\cite{danijuan}. By solving the Schr\"odinger equation in momentum space in coupled channels and normalizing the wave function of the bound state to unity, it was found \cite{danijuan} 
\begin{equation}
\int d^3p \mid \langle p \mid \Psi_i \rangle \mid^2 = g_i^2 \frac{\partial G_i}{\partial E} \ ,
\end{equation}
with $\mid \Psi_i \rangle$ being the $i$ component of the bound state in the $i^{\rm th}$ channel, so that each term of the sum in Eq.~(\ref{g}) represents the probability to have this channel in the wave function of the bound state:\footnote{As discussed in \cite{danijuan} there is a different normalization of the amplitudes, and hence the couplings, between \cite{danijuan} and field theoretical approach used here, which leaves the probability to be expressed here as in Eq.~(\ref{P})}
\begin{equation}
\label{P}
P_i=-g_i^2\frac{\partial G_i}{\partial s} \ ,
\end{equation}
and the sum of these probabilities saturates the wave function. 
Note that, by construction, in the case we are discussing here all the components of the composite state are of meson-meson type. We will elaborate more on these issues in Sect.~\ref{systematic}.

It is easy to visualize a genuine state that couples weakly to a meson-meson component by using a meson-meson potential of the type:
\begin{align}
V=\frac{b}{s-s_R} \ ,\label{CDD}
\end{align}
which we refer to as a CDD pole~\cite{castillejo}. Now
\begin{align}
T=\frac{1}{\dfrac{s-s_R}{b}-G},\quad g^2=\frac{1}{\dfrac{1}{b}-\dfrac{\partial G}{\partial s}},
\end{align}
and 
\begin{align}
P=-g^2\frac{\partial G}{\partial s}=1-g^2\frac{1}{b}.
\end{align}
In the limit of $b\to 0$ (small coupling of the genuine state to meson-meson) we have $g^2\to 0$ and the pole appears at $s=s_R$. Then the amount of meson-meson component, $-g^2 \partial G/ \partial s$, goes to zero and we have a representation for a genuine state, or, in general, a state different from the explicit two meson state considered.
It is interesting to note a distinct feature in the potential of Eq.~(\ref{CDD}), namely its energy dependence.

These ideas are generalized in Ref.~\cite{hyodoreport}, with the sum rule 
\begin{align}
-\sum_i g^2_i\frac{\partial G_i}{\partial s}-\sum_{i,j}g_i g_j G_i\frac{\partial V_{i,j}}{\partial s} G_j=1,\label{Hy}
\end{align}
evaluated at the pole. The first term in Eq.~(\ref{Hy}) is associated in Ref.~\cite{hyodoreport} to the composite part of the state (meson-meson in the present case) and the second term, involving the derivative of the potential, to the genuine part of the state. Actually, this second part accounts for the state components that have not been considered in the coupled channel problem. This is easily shown in the case of two channels in Ref.~\cite{acetidelta}, where one channel is eliminated and its effects are accounted for by means of an effective potential in the remaining channel. Take $V_{22}=0$, for simplicity, and consider $V_{ij}$ energy independent to saturate the state with the two channels in Eq.~(\ref{V2ch}). It is then easy to obtain from Eq.~(\ref{T2ch}),
\begin{align}
T_{11}=\frac{V_{11}+V^2_{12}G_2}{1-(V_{11}+V^2_{12}G_2)G_1} \ ,
\end{align}
making clear that solving  a one-channel problem with the effective potential 
\begin{align}
V_\textrm{eff}=V_{11}+V^2_{12}G_2\ ,
\end{align}
gives the  same amplitude $T_{11}$ obtained in the two channel case. The novelty is that now $V_\textrm{eff}$ becomes energy dependent. Then, the term $-g^2_1 \partial G_1/\partial s$, which accounts for the probability of channel 1 in the state, is the same in both formulations and the second term in Eq.~(\ref{Hy}) is, by construction of $V_\textrm{eff}$, the probability of the second channel that has been eliminated.   We are going to use these findings to analyze the lattice spectra of Ref.~\cite{sasa}.

\section{Analysis of the lattice spectra}

The lattice simulation of Ref.~\cite{sasa} obtained three energy levels in the scalar channel using the $K D$ and $\bar sc$ interpolators, and three\footnote{{The second level in the axial channel of Ref.~\cite{sasa} is attributed to the $D_{s1}(2536)$ resonance in $K D^*$ d-wave scattering and is therefore not used in the present paper which considers $K D^*$ scattering in s-wave. In principle  $L=0$ and $L=2$ can mix for $J=1$, but using arguments of Heavy Quark Spin Symmetry \cite{isgurwise}, the spin of the heavy quark $\vec{S}_Q$ is conserved, and so is $\vec{J}$, and
hence $\vec{J}-\vec{S}_Q$, which can be constructed from $\vec{L}$ and the spin 1/2 of the light quark of the $D^*$. For $L=0$,  $\vec{J}-\vec{S}_Q$ only has modulus $1/2$, and for $L=2$, it can have the values 3/2 and 5/2. Thus, $L=0$ and $L=2$ do not mix at leading order in the $O(1/m_Q)$ expansion. }} levels in the axial channel using the $K D^*$ and $\bar sc$ interpolators.  Table \ref{En} collects the levels of ensemble (2)\footnote{{Results of set 2 in \cite{sasa} are used in the axial channel.}}, {with $N_f\!=\!2+1$ and close-to-physical pion mass $m_\pi =156$ MeV.   The lattice spacing is $a=0.0907\,(13)$ fm and the box size $L=2.90$ fm. 
The kaon with mass $m_K=504(1)~$MeV obeys the usual  relativistic dispersion relation   $E_K(p)=(m_K^2+p^2)^{1/2}$. }

\begin{table}[ht]
\begin{center}
\caption{Energy levels for the scalar ($K D$)   and   axial ($K D^*$) channels found in the simulation Ref.~\cite{sasa}.     The relative errors in the lattice spacing $a$ and in $a\,E$ have been added in quadrature. Only the energy differences, for example $E_n^{\rm lat}-\bar m_{D_s}^{\rm lat}$ with $ \bar m_{D_s}^{\rm lat}=\tfrac{1}{4}(m_{D_s}+4m_{D_s^*})=1.8407(6)~$MeV,  can be compared to the experiment. }\label{En}
\begin{tabular}{c|c|c}
\hline
&$KD$ channel &$KD^*$ channel\\
\hline
$E_1$ (MeV)&2086 (34) &2232 (33)\\
$E_2$ (MeV)&2218 (33) &2349 (34)\\
$E_3$ (MeV)&2419 (36) &2528 (53)\\
\hline
\end{tabular}
\end{center}
\label{table}
\end{table}

{The simulation \cite{sasa,sasa1} treated the charm quark using the so-called Fermilab method, where the leading discretization errors related to the charm quark cancel in the energy differences (with respect to the reference mass of a meson containing the same number of charm quarks). 
We employ the dispersion $E(p)$ for $D$ and $D^*$ mesons determined in the simulation of Ref.~\cite{sasa}}   
\begin{align}
E_{D(D^*)} (\vec{p}\,)=M_1+\frac{\vec{p}^{\,2}}{2M_2}-\frac{(\vec{p}^{\,2})^2}{8M^3_4}\ ,\quad m_{D(D^*)}=M_1\label{disp}
\end{align}
where $M_1$, $M_2$, $M_4$ are given in Table~\ref{tableM}. 
\begin{table}[ht]
\begin{center}
\caption{$M_i$ from the  dispersion relation $E(p)$ (\ref{disp}) for $D$ and $D^*$ mesons. The rest energies, i.e. the masses $M_1$, can be compared to experiment via the difference $M_1^{\rm lat}-\bar m_{D}^{\rm lat}$ with $ \bar m_{D}^{lat}=\tfrac{1}{4}(m_{D}+4m_{D^*})=1.751(3)~$MeV \cite{sasa}. } \label{tableM}
\begin{tabular}{c|c|c}
\hline
& {$D$ meson} & {$D^*$ meson}\\
\hline
$M_1$ (MeV)&1639&1788\\
$M_2$ (MeV)&1801&1969\\
$M_4$ (MeV)&1936&2132\\
\hline
\end{tabular}
\end{center}

\end{table}

\subsection{Analysis by means of the effective range formula}

In Ref.~\cite{sasa} the scattering length and effective range for $KD$ and $KD^*$ scattering were obtained using only the two lowest energy levels of the lattice simulation and employing L\"uscher's approach to {extract the infinite volume phase shifts}. In this section we analyze these results by means of an effective range formula to obtain the binding energy of the state and check the fulfillment of the sum-rule of  Eq.~(\ref{LH}).

The effective range approximation reads
\begin{align}
 p\,\cot\delta=\frac{1}{a_0}+\frac{1}{2}r_0 p^2, \quad T=-\frac{8\pi E}{p\,\cot\delta-{\rm i}p}.\label{range}
\end{align}
Below threshold, one writes $p={\rm i}\tilde{p}$, and a pole of the $T$ matrix is obtained for  $\cot\delta={\rm i}$. 
Therefore, the pole appears for the value of $\tilde p$ that satisfies 
\begin{align}
\frac{1}{2}r_0\tilde{p}^2-\tilde{p}-\frac{1}{a_0}=0. 
\end{align}
Taking random $a_0$ and  $r_0$ values within the range determined by the lattice simulation \cite{sasa}, quoted in Table \ref{tab:effective}, we obtain a series of values for the bound momentum $\tilde{p}$ and the corresponding binding energy
\begin{align}
B=-\frac{\tilde{p}^2}{2\mu},\quad \mu=\frac{m_K m_{D/D^*}}{m_K+m_{D/D^*}} \ .
\end{align}
The average value of the binding energy for the $KD$ state, which is associated to the $D^*_{s0}(2317)$, is then found to be 38(9) MeV. We note that the unitary coupled-channel approach of \cite{dani} generates such a state from the interaction of the $KD$ and $\eta D_s$ channels mostly. Had we used  the central values of $a_0$ and $r_0$ directly, we would have obtained $B=35.8$ MeV, which obviously lies within the error bar of the results quoted in Table \ref{tab:effective}. We note that this value is 0.8 MeV smaller than the one given in \cite{sasa}, essentially because in the present analysis we have used the (isospin averaged) physical masses of the mesons instead of the lattice ones. Employing the same procedure, we find a $K D^*$ state with a binding energy of 44(6) MeV, which we associate to the $D_{s1}^*(2460)$. In the unitary coupled-channel approach this state is mainly built from $KD^*$ and $\eta D^*_s$ components \cite{daniaxial}. 

\begin{table}[h]
  \setlength{\tabcolsep}{0.3cm}
\begin{center}
\begin{tabular}{|l|c|c|c|c|c|}
\hline
{Channel}  & $a_0$ [fm] & $r_0$  [fm] & $B$ [MeV] & $|g|$ [GeV] & $-g^2\partial G/\partial s$ \\
\hline
$KD$ & $-1.33(20)$ & $0.27(17)$ & 38(9) & $12.6(1.5)$ & 1.14(0.15) \\
\hline 
$K D^*$ &  $-1.11(0.11)$ & $0.10(0.10)$ & $44(6)$ & $12.6(0.7)$ & $0.96(0.06)$ 
\\
\hline
\end{tabular}
\caption{Binding energy $B$, meson-meson coupling $|g|$ and sum-rule [Eq.~(\ref{LH})], for the bound states obtained in the lattice QCD simulation of
Ref.~\cite{sasa}, analyzed using an effective range formula. }\label{tab:effective}
\end{center}

\end{table}

It is interesting to test the sum rule of Eq.~(\ref{LH}) for the states obtained. The $g^2$ at the pole can be expressed as   
\begin{align}
g^2=\frac{16\pi s\tilde{p}}{\mu (1-r_0\tilde{p})} \ ,
\end{align}
 and listed in Table \ref{tab:effective}.
Since $\partial G/\partial s$ is convergent, we obtain the sum rules quoted in the last column, which, within errors,
are all compatible with unity.  The coupling to the $KD$ channel is  $g_{KD} = 12.6$ GeV, which is of the order of the one obtained in the chiral unitary approach in Ref.~\cite{dani}, $g_{KD}=10.21$ GeV. Note, however, that this smaller value would provide a probability for the $KD$ channel of about 
$60-70\%$, leaving room for the other channels considered in the unitary coupled-channel approach. Similarly, in the   $KD^*$ channel, we find a coupling $g_{KD^*} =  12.6~$GeV,  compared  to the value of around 10 GeV quoted in Ref.~\cite{daniaxial}, also leaving room for the additional meson-meson components considered in that work.

Although  the results obtained with the effective range formula are qualitatively reasonable, and the existence of the bound state emerges as a solid statement, one can see that the approximation has its limitations when one looks at other magnitudes like the probability $P(KD)$, which comes out larger than one (although compatible within errors). There is also the fact that the first two levels are separated by 132 MeV, which makes this approximation a bit extreme.  Furthermore, the information of the third level is not used, and, as shown in Ref.~\cite{sasa}, this level cannot be accounted for by means of the effective range formula. All these reasons advise a new reanalysis which we offer in the next subsection.

\subsection{Analysis of lattice spectra by means of an auxiliary potential}
\label{aux}
First, we are going to make the analysis with only one channel. Anticipating that the $\eta D_s$ and $\eta D^*_s$ channels also play a role in the $D^*_{s 0} (2317)$ and $D_{s1}(2460)$ resonances, as found in Refs.~\cite{dani} and
~\cite{daniaxial}, we shall leave room for these and possible $\bar qq$ components, by using an energy dependent potential. As a first step we take a potential linear in $s$, 
\begin{align}
V=\alpha+\beta(s-s_{\rm th}),\label{pot}
\end{align}
with $s_{\rm th}=(M_{D^{(*)}}+M_K)^2$,
since only the derivative of the potential is needed to obtain the sum rule. Later on we shall also use another type of potential.

 In the finite box, the $T$ matrix of Eqs.~(\ref{BS}) and~(\ref{T}) is replaced by
 \begin{align}
 \tilde{T}=\frac{1}{V^{-1}-\tilde{G}},\label{Tfin}
 \end{align}
where $\tilde{G}$ is the two meson loop function in the box given by~\cite{koren}
\begin{equation}
\tilde{G}=G+\lim_{q_\textrm{max}\to\infty}\left[\frac{1}{L^3}\sum_{q_i}^{q_\textrm{max}}I(\vec{q}_i)-\int\limits^{}_{q<q_\textrm{max}}\frac{d^3q}{(2\pi)^3}I(\vec{q}\,)\right]\ ; \hspace{1cm} \vec{q}_i = \frac{2\pi}{L}\vec{n}_i, ~~\vec{n}_i \in \mathbb{Z}^3\ .
\end{equation}
The $G$ in the continuum, Eq.~(\ref{eq:loop}), can be regularized with a cut-off $q^\prime_\textrm{max}$ or employing dimensional regularization. The latter choice, followed in Ref.~\cite{koren}, cannot be applied here because we employ the dispersion relation of Eq.~(\ref{disp}). For this reason we adopt the cut-off method, with a cut-off value that gives equivalent results to those of the chiral unitary approach of Refs.~\cite{dani,daniaxial}. Any value {of $q_{\rm max}'$} can, in principle, be taken since changes in $G$ can be accommodated by changes in $V^{-1}$ when we require that $\tilde{T}$ has poles at the energies of the lattice spectra by demanding that $V^{-1}-\tilde{G}=0$. Note, in addition, that we are interested finally in results for the continuum. Hence, at the energies of the lattice spectra we have $V^{-1}=\tilde{G}$, and then the continuum $T$ matrix is
\begin{equation}
T=\frac{1}{V^{-1}-G}=\frac{1}{\tilde{G}-G}=\frac{1}{\displaystyle\lim_{q_\textrm{max}\to\infty}\left[\frac{1}{L^3}\displaystyle\sum_{q_i}^{q_\textrm{max}}I(\vec{q}_i)-
\displaystyle\int\limits^{}_{q<q_\textrm{max}}\displaystyle\frac{d^3q}{(2\pi)^3}I(\vec{q}\,)\right]},\label{Tlim}
\end{equation}
which is  then independent of the cut-off $q^\prime_\textrm{max}$ employed to regularize $G$. However, in the transfer of strength from $G$ to $V^{-1}$ one will be introducing some energy dependence in $V^{-1}$ that would change the probability $Z$ of not having the main meson-meson component considered. We shall come back to this issue in section \ref{systematic} where systematic uncertainties are studied. 

Equation (\ref{Tlim}) is the formulation employed in the approach of Ref.~\cite{misha}, where it is shown  that L\"uscher formula is recovered if some terms of $I(\vec{q}\,)$, which are exponentially suppressed, are eliminated. These terms can be relevant in the case of relativistic particles and small volumes ~\cite{chen,chennew}, which is not the case here.  However, we cannot use the standard L\"uscher approach either, based on the relativistic relationship $\omega (q)=(m^2+q^2)^{1/2}$, since we are forced to employ the dispersion relation of Eq.~(\ref{disp}). In this case,   Eq.~(\ref{Tlim}) gives the appropriate extension of the L\"uscher formalism.

 There is another approximation inherent in our approach (or the one of L\"uscher) when we assume that the potential is volume independent. Within the framework of the chiral unitary approach such effects were investigated in \cite{miguerios,luismigue} in the $\pi \pi$ scattering in the scalar sector and the $\rho$ sector and it was concluded that for values of $L m_{\pi}> 1.5$ they could be safely neglected. In the present case, given the large masses involved, loops in the t-channel, which originate this volume dependence, are even less relevant. 

 With the formalism exposed above, a best fit is carried to the three lattice levels obtained in~\cite{sasa}, demanding that the $\tilde{T}$ derived from Eq.~(\ref{Tfin}) using the potential of Eq.~(\ref{pot}) has poles at the three energies. In order to find the desired magnitudes and associated statistical errors, we perform a series of fits to different sets of three energies, generated with a Gaussian weight within the errors of the lattice levels. With the parameters obtained in each fit we evaluate the different magnitudes. From the results obtained in the different fits, we then determine the central values and statistical errors of these magnitudes.

\begin{figure} 
\begin{center}
\includegraphics[width=0.5\textwidth,clip]{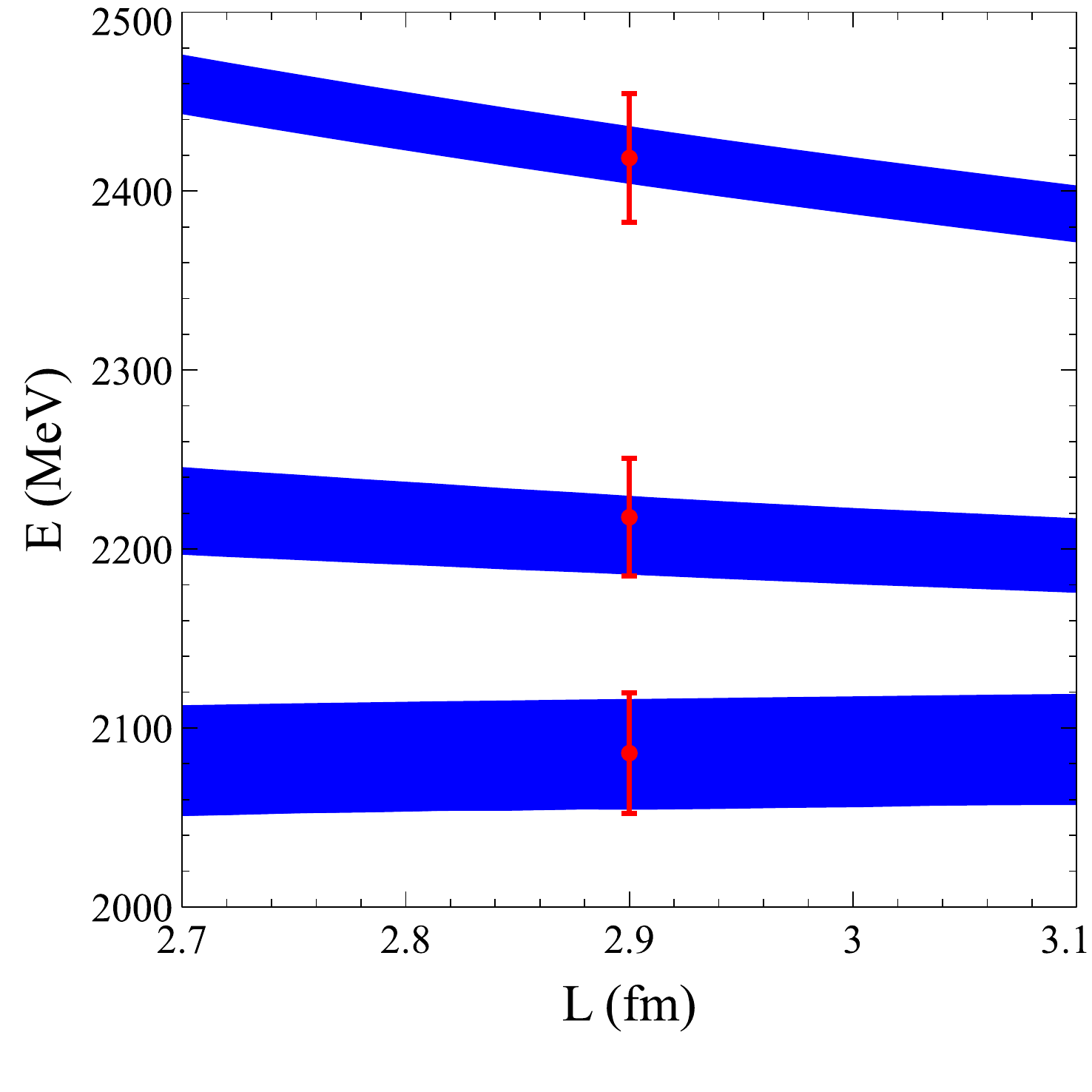}
\caption{Fits to the lattice data of Ref.~\cite{sasa} for the $KD$ system using the potential of Eq.~(\ref{pot}).}\label{fDK1}
\end{center}

\end{figure}

\begin{figure}
\begin{center}
\includegraphics[width=0.5\textwidth,clip]{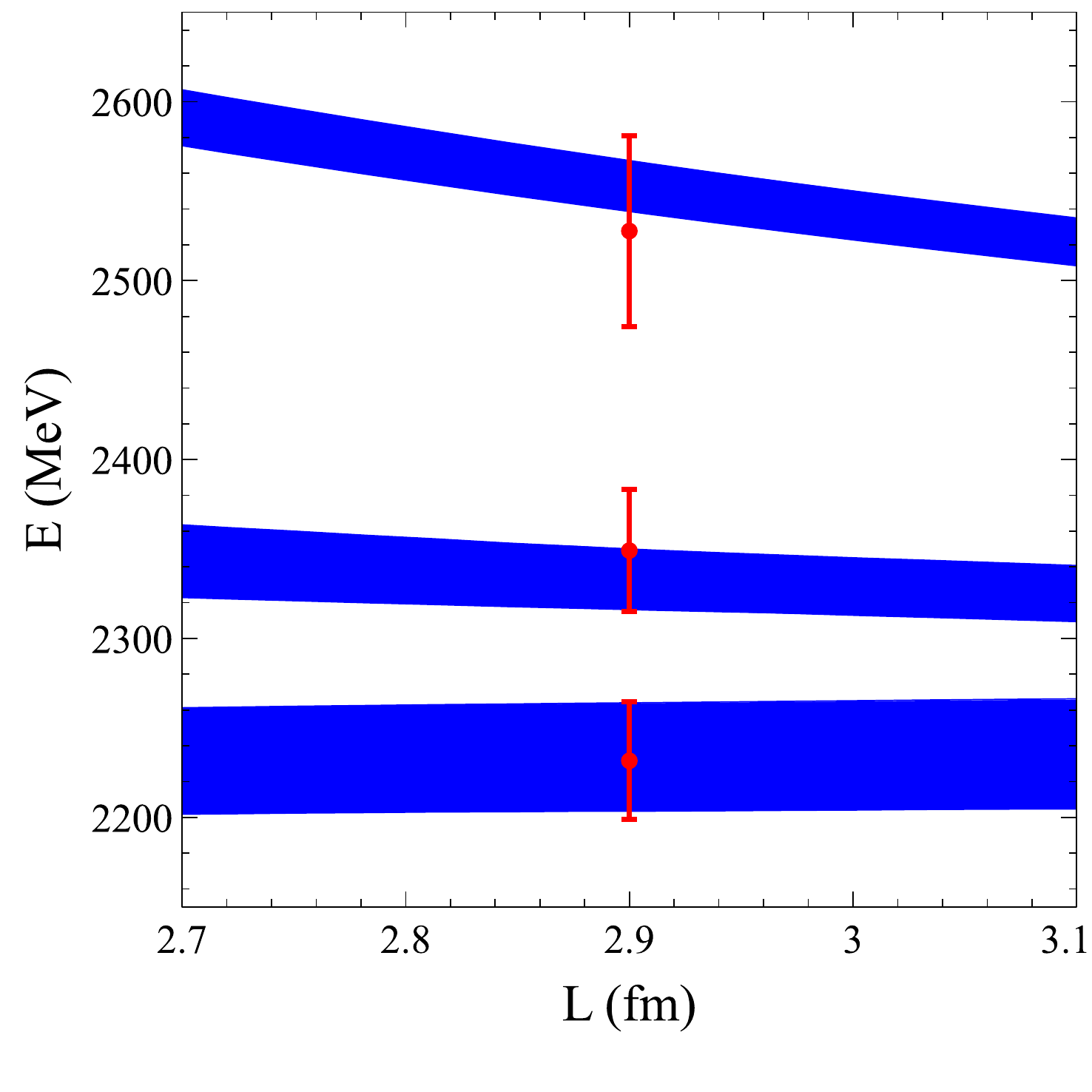}
\caption{Fits to the lattice data of Ref.~\cite{sasa} for the $KD^*$ system using the potential of Eq.~(\ref{pot}).}\label{fDstarK1}
\end{center}

\end{figure}

We show in Figs.~\ref{fDK1} and~\ref{fDstarK1} the results obtained from the fits to the levels for the $KD$ and $KD^*$ systems, respectively.
The procedure outlined above gives us a pole for the $KD$ system  with  binding energy
\begin{align} 
B(KD)=m_D+m_K-E_B(K D)=46\pm 21~\textrm{MeV}\ ,
\label{binding}
\end{align}
to be compared to the value 36.6(16.6)(0.5) MeV obtained with the effective range formula in \cite{sasa,sasa1} and to the 45 MeV binding in the physical case.
For the $KD^*$ system we get the binding energy
\begin{align} 
B(KD^*)=m_{D^*}+m_K-E_B(K D^*)=52\pm 22~\textrm{MeV}\ .
\label{binding1}
\end{align}

The probabilities for the $KD$, $KD^*$ components, obtained from Eqs.~(\ref{P}), (\ref{gsq}), are:
\begin{align}
P(KD)&=(76\pm 12)~\%,~\textrm{for the $D^*_{s0}(2317)$}\ ,\label{primone}\\
P(KD^*)&=(53\pm 17)~\%,~\textrm{for the $D_{s1}(2460)$}\ .\label{P1ch}
\end{align}
This means that there is a large amount of $KD$ and $KD^*$ components in the corresponding bound states. 

\subsection{Fit with a CDD pole}
\label{sec:cdd}
{ One near-threshold level was found in \cite{sasa,sasa1} when only $\bar sc $  interpolators were used\footnote{Its energy however changes when $D^{(*)}K$ interpolators were used in addition to $\bar sc$ ones.}, and one wonders what is the $\bar sc$ component in the meson states at hand.} We therefore explore whether there could be an admixture of some genuine component in the bound state by refitting the lattice levels adding a CDD pole to the potential of Eq.~(\ref{pot}):
\begin{equation}
V=\alpha+\beta(s-s_{\rm th})+ \frac{\gamma^2}{s-M_{\rm CDD}^2} \ ,
\label{eqcdd}
\end{equation}
which, as seen in section \ref{section2}, is suited to accommodate a genuine state. This has been shown to be the proper way to account for genuine components in different works \cite{acetidelta,nsd,aceti_oset,xiao_aceti_bayar} in the continuum.
An analysis of  ``synthetic'' lattice spectra in terms of this potential was done in
\cite{koren}. It was also recently employed to analyze lattice spectra with the $\pi K$ and $\eta K$ channels in \cite{Dudek:2014qha}.

Since we have four parameters ($\alpha$, $\beta$, $\gamma$ and $M_{\rm CDD}$) and three energy levels, we can obtain solutions with many sets of parameters which are, obviously, correlated. However, the values of the parameters do not have a particular significance and what matters is the value of the magnitudes derived from the different fits. The statistics of the obtained fits shows a clear preference for solutions with a $M_{\rm CDD}$ value that lies far away (more than 300 MeV)  from the  $KD$, $KD^*$ thresholds, such that it effectively provides a linear dependence in $(s-s_{\rm th})$ at the energies where the poles are found. This is an indication that the lattice energies do not favour a CDD component, or at least not a significant one. Obviously, future lattice results with more accuracy  and different volumes will allow one to be more precise on this issue.

   With the potential of Eq.~(\ref{eqcdd}) we obtain the following binding energies
\begin{align}
B(KD)=29\pm 15~\textrm{MeV}\ , \label{binding2}\\
B(KD^*)=37\pm 23~\textrm{MeV}\ \label{binding3},
\end{align}
and probabilities
\begin{align}
P(KD)&=(67\pm 14)~\%,~\textrm{for the $D^*_{s0}(2317)$}\ ,\label{primetwo}\\
P(KD^*)&=(61\pm 26)~\%,~\textrm{for the $D_{s1}(2460)$}\ ,\label{P1chtwo}
\end{align}
which are compatible within errors with those of Eqs. (\ref{binding})--(\ref{P1ch}),  obtained with the linear potential.

\subsection{Two channel analysis}

After this exercise we perform a two channel analysis including the $\eta D_s$ channel for the $D^*_{s0}(2317)$ state and the $\eta D^*_s$ channel for the $D_{s1}(2460)$, which were found also relevant in Refs.~\cite{dani,daniaxial}.

Since we only have three energy levels we use an energy independent potential, Eq.~(\ref{V2ch}), which has three parameters, $V_{11},V_{12},V_{22}$. By doing so, we would force the states to saturate with the $KD^{(*)}$, $\eta D_s^{(*)}$   components. The comparison of the two procedures would allow us to make statements about the amount of each channel in the respective states.

We thus fit the $V_{ij}$ parameters using
\begin{align}
\tilde{T}=(1-V\tilde{G})^{-1} V,
\end{align}
in two channels, looking for the poles of $\tilde{T}$ and associating the first three levels to those of the lattice simulation.  

We do not find any suitable fit to the data, which is an enlightening result. One could interpret it as an evidence that the energy levels obtained in \cite{sasa} do not contain information on the $\eta D_s$ or $\eta D^*_s$ channels. This seems to be the case because the three energies obtained there were tied to the use of $q \bar q$ and meson-meson interpolators of $KD$ or $KD^*$ type. No interpolator was used containing information on the $\eta D_s$ and $\eta D^*_s$ channels, and no energy level was found which would be tied to these channels. 
It is indeed a common experience of  lattice practitioners that a given two-hadron eigenstate is most often not seen unless explicitly implemented in the basis of interpolating fields. Although all states with a given quantum number are in principle expected in a dynamical simulation, a poor basis of interpolating fields is insufficient to render them in practice. The reason is that one would have to wait much time till these components show up in the time evolution of the state and this could happen in the region where the ratio of noise to signal is large, preventing any signal to be seen \cite{wittig}. 
This also gives us some idea on how to proceed in the future if one wishes to make progress on determining the components of the $D^*_{s0}(2317)$ and $D_{s1}(2460)$ wave functions. The relevant fraction of the wave function that went to the $\eta D_s$ and $\eta D^*_s$  channels in chiral unitary studies \cite{dani,daniaxial}, of the order of 20\%, makes it advisable to include interpolators for the  $\eta D_s$ and $\eta D^*_s$ channels in future lattice simulations. {Such a simulation is underway and preliminary spectra have been presented in \cite{ Ryan:lat14}. }. In any case, it is worth stressing that, as shown in previous sections, the present lattice information allows to conclude that there are extra components to the dominant $KD$ in the $D_{s0}^*(2317)$ wave function, although one cannot state which ones.

\section{Scattering length and effective range}
We can also obtain the scattering length and the effective range in each of the cases explored. For this we use Eq.~(\ref{range}), finding 
\begin{align} 
p\,\cot\delta= \textrm{Re}\left\{-\frac{8\pi E}{T}\right\} \simeq \frac{1}{a_0}+\frac{1}{2}r_0 p^2~.
\label{efferange}
\end{align}
Relating $E$ to $p$ via the dispersion relation of Eq.~(\ref{disp})
\begin{align}
E=\sqrt{m^2_K+p^2}+E_{D(D^*)}(p) \ ,
\end{align}
we obtain
\begin{align}
a_0&=-1.2\pm 0.6~\textrm{fm},\quad r_0=0.04\pm 0.16~\textrm{fm}~\textrm{for $KD$},\\
a_0&=-0.9\pm 0.3~\textrm{fm},\quad r_0=-0.3\pm 0.4~\textrm{fm}~\textrm{for $KD^*$}
\end{align}
in the case the lattice data is analyzed using a single channel potential (\ref{pot}).

When we use the CDD potential of Eq. (\ref{eqcdd}) we find 
\begin{align}
a_0&=-1.4\pm 0.4~\textrm{fm},\quad r_0=-0.2\pm 0.4~\textrm{fm}~\textrm{for $KD$},\\
a_0&=-1.3\pm 0.6~\textrm{fm},\quad r_0=-0.1\pm 0.2~\textrm{fm}~\textrm{for $KD^*$}
\end{align}
 The values for the scattering length and effective range obtained with the different methods are remarkably similar.

The values obtained also agree qualitatively with those obtained in Ref.~\cite{sasa}. Yet, as we have discussed, we do not use the effective range formula to correlate the results. Indeed, instead of Eq. (\ref{efferange}) we have 
\begin{align} 
p\,\cot\delta= -8\pi E(V^{-1}- \textrm{Re} \{G\})
\label{neweq}
\end{align}
and the $G$ function depends on the cut off. If $V$ is energy independent we have two degrees of freedom in the approach to accommodate the values of $a_0$ and $r_0$, but $p\,\cot\delta$ develops terms in $p^4$ which are tied to the values of  $a_0$ and $r_0$. If we allow $V$ to be energy dependent, as in Eq. (\ref{pot}), we have more freedom to accommodate the $p^4$ terms in the expansion of $p\,\cot\delta$. However, the main problem in the use of Eq. (\ref{efferange}) is that it blows up at large energies, where the series expansion does not converge. Our method, which does not make a series expansion of $p\,\cot\delta$, has a good behavior at higher energies from the analytical behavior of $\textrm{Re} \{G\}$, which contains the $log$ terms of the intermediate particle propagators.
This allows us to cover a wider span of energies and we can make use of the three energy levels obtained in \cite{sasa}, while only the information of the lowest two could be accommodated in the analysis of \cite{sasa} based on Eq. (\ref{efferange}). 

\section{Evaluation of systematic uncertainties}
\label{systematic}

In \cite{hanhart_orginos} the lowest lattice level obtained for the channels $D \bar K (I=1)$, $D \bar K (I=0)$, $D_s K$, $D \pi (I=3/2)$, $D_s \pi$, free from disconnected diagrams, were employed to obtain, via the L\"uscher formalism \cite{composite},  the phase shifts in the continuum at the eigenenergies of the lattice box. The scattering length was then derived from the relationship  
$p\,\cot\delta(p)= 1/a_0$, 
disregarding the effective range term. The low energy constants of a chiral lagrangian were fitted to the scattering lengths of those channels employing a unitary approach. With these values of the coefficients, the coupled $KD$, $\eta D_s$ channels system was studied, from where the existence of a bound state associated to the $D_{s0}^*(2317)$ was established and the $KD$ scattering length was obtained. A $KD$ probability, $1-Z$, in the $D_{s0}^*(2317)$ wave function of around 70\% was found, where the value of $Z$ was determined from the scattering length via the relation \cite{composite,baru}
\begin{equation}
a_0= -2 \frac{(1-Z)}{(2-Z) } \frac{1}{\sqrt{ 2\mu \epsilon}} \left[1+ {\cal O}\left(\sqrt {2\mu \epsilon}/ \beta\right)\right] \ ,
\label{eq:a0_weinberg}
\end{equation}    
with $\mu$ and $\epsilon$ being the reduced mass and binding energy, respectively, and $1/\beta$ accounting basically for the range of the interaction ($1/q_{\rm max}$ in our approach). The term ${\cal O}(\sqrt {2\mu \epsilon}/ \beta)$, negligible for small binding energies, is often discussed as uncertainty.  
In the present case $\sqrt {2\mu \epsilon}/ \beta$ is of the order of 0.22 if we take $\beta=q_{\rm max}=M_V=780$~MeV, and the correcting terms can be relevant. 

Indeed, 
let us comment on the sensitivity of Eq.~(\ref{eq:a0_weinberg}) in obtaining $Z$ from the value of $a_0$. Note that if $-2/\sqrt{2\mu \epsilon} < a_0 < -1/\sqrt {2\mu \epsilon}$, the resulting $Z$ would have unphysical negative values. This condition would obviously not be a problem for sufficiently small binding energies where Eq.~(\ref{eq:a0_weinberg}) is applicable but, for the $KD$ state analyzed here, the value of the factor  $-1/\sqrt {2\mu \epsilon}$ is $-1.12$ fm, close to the typical values found for the scattering lengths, and this can lead to large uncertainties in the extraction of $Z$ from $a_0$ using Eq.~(\ref{eq:a0_weinberg}).  Note that Ref.~\cite{hanhart_orginos} obtained $a_0\sim -0.85$ fm, from which, using  Eq.~(\ref{eq:a0_weinberg}), a probability $P_{KD}\sim 70\%$ was extracted, similar to the result obtained here in spite of the fact that we have a different value of the scattering length. 

Incidentally, one could have evaluated $P=1-Z$ directly from the coupling also in the Weinberg approach using Eq.~(24) from Ref.~\cite{composite}, which is equivalent to Eq.~(\ref{P}) used here but neglecting the ${\cal O}(\sqrt {2\mu \epsilon}/ q_{\rm max})$ terms in $(\partial  G/ \partial s)$ and in the determination of $g_i^2$. It is instructive to see the correcting terms in $(\partial  G/ \partial s)$ due to the range of the interaction. Using, for simplicity,  the nonrelativistic approach of \cite{danijuan} (see Eqns. (27), (29) there) one finds
\begin{eqnarray}
\frac{\partial G}{\partial E} &=& \frac{1}{\gamma} 8\pi \mu^2\left[\arctan\left(\frac{q_{\rm max}}{\gamma}\right) - \frac{\gamma q_{\rm max}}{\gamma^2+ q_{\rm max}^2}\right] = \\
&=& \frac{1}{\gamma} 8\pi \mu^2\left[ \frac{\pi}{2} - 2\left(\frac{\gamma}{q_{\rm max}}  \right) + \frac{4}{3} \left(\frac{\gamma}{q_{\rm max}} \right)^3 + \dots \right] = \\
&=& \frac{1}{\gamma} 4\pi^2 \mu^2\left[ 1 - \frac{4}{\pi}\left(\frac{\gamma}{q_{\rm max}}  \right) + \frac{8}{3 \pi} \left(\frac{\gamma}{q_{\rm max}} \right)^3 + \dots \right]  \ .
\label{dGnonrel}
\end{eqnarray}
Hence, in the nonrelativistic expression
\begin{equation}
1-Z= g^2 \frac {\partial  G} {\partial E}\ ,
\label{Pnr}
\end{equation}
analogous to Eq.~(\ref{P}),
the correcting factor to the Weinberg formula from range effects in $\partial  G/\partial E$ is\footnote{The normalizations for $g$ in [12] and here are different. In [12], or in the Weinberg notation, $\partial G/\partial E$ is used instead of $\partial G/\partial s$, but the range correcting factor, $F$, is the same.}:
\begin{eqnarray}
F= \left[ 1 - \frac{4}{\pi}\left(\frac{\gamma}{q_{\rm max}}  \right) + \frac{8}{3 \pi} \left(\frac{\gamma}{q_{\rm max}} \right)^3 + \dots \right]  \ .
\label{correfac}
\end{eqnarray}
to which one would have to add the correcting terms to the expression of $g^2$ in Ref. \cite{composite}. 
The deviation from unity of Eq. (\ref{correfac}) in the problem analyzed here amounts to 28\%. Although one would also have correcting terms from $g^2$, this exercise gives us an idea of the order of magnitude of the corrections due to finite range effects in the determination of $1-Z$. The exercise also serves us another purpose, which is to note that employing Eq. (24) from Ref. \cite{composite} can give reasonable numbers for $1-Z$ in the present case, within uncertainties, while applying Eq. (\ref{eq:a0_weinberg}) is not possible for a value $a_0\sim -1.3$ fm. Actually, in Ref. \cite{guonew}, following the work of \cite{hanhart_orginos}, the value of $a_0\sim -1.33$ fm from the lattice work of \cite{sasa1} is used as input to further constrain the parameters of the chiral theory, but Eq. (\ref{eq:a0_weinberg}) is no longer used. In our case we do not use Eq. (\ref{eq:a0_weinberg}), nor Eq. (24) from Ref. \cite{composite}, in order to determine $1-Z$, but Eq. (\ref{P}) in which explicit range effects will appear in $g^2$ and  $\partial G_i/\partial s$ from our formulation of the problem using Eq. (\ref{T}) for the scattering of the particles. This is our prescription to take into account range effects and we discuss next the sensitivity of the results to the changes of the range parameter $q_\textrm{max}$, also within our approach.

We estimate the uncertainties inherent to the method for not too small binding energies, like in the present case, by performing fits to the lattice energies employing four different values of $q_{\rm max}$, 770 MeV, 875 MeV, 1075 MeV, 1275 MeV, and the auxiliary potential linear in $s$ of Eq.~(\ref{pot}). This will inform us on the size of systematic uncertainties coming from this source. In order not to be confused by the statistical uncertainties, the fit for each value of $q_{\rm max}$ will be done to the central values of the lattice energies. Our results, shown in Table \ref{qmax_1} for the $KD$ system and in Table \ref{qmax_2} for the $K D^*$ one, confirm that the systematic uncertainties tied to the range are small and  well within the statistical uncertainties. The binding energy of the $K D^*$ system shows a stronger sensitivity to the heavy meson mass employed than that of all other magnitudes, the changes of which fall well within the statistical errors.

\begin{table}[ht]
\caption{Dependence of the properties of the $KD$ bound state on  $q_{\rm max}$}\label{qmax_1} 
\begin{center}
\begin{tabular}{l|c|c|c|c|c}
\hline
$q_{\rm max}$ (MeV) & 770& 875 &  1075 & 1275 & Average\\
\hline
$B $(MeV) & 34.2& 36.6 & 35.5 & 35.5 & $35.5\pm0.8$ \\
$\mid g\mid$ (GeV) & 10.85&10.60 & 10.37 & 10.41 & $10.6\pm 0.20$ \\
$P$ (\%)  & 86.68& 82.15 &  84.09 & 87.16 & $85\pm2$ \\
$a_0$ (fm) & $-1.32$&$-1.24$ & $-1.25$ & $-1.25$ &  $-1.27\pm0.03$ \\
$r_0$ (fm) & 0.30&$0.22$ & $0.19$ & $0.19$  & $0.23\pm0.05$ \\
\hline
\end{tabular}
\end{center}
\end{table}

\begin{table}[ht]
\caption{Dependence of the properties of the $KD^*$ bound state on  $q_{\rm max}$}\label{qmax_2} 
\begin{center}
\begin{tabular}{l|c|c|c|c|c}
\hline
$q_{\rm max}$ (MeV) & 770&875 &  1075 & 1275 & Average\\
\hline
$B$ (MeV) & 45.8&45.6 & 44.9 & 44.2 & $45.0\pm 0.7$ \\
$\mid g\mid$ (GeV) & 10.67&10.15 & 10.32 & 10.31 & $10.4 \pm 0.2$ \\
$P$ (\%)  & 60.30&57.42 &  63.33 & 66.10 & $62\pm3$ \\
$a_0$ (fm) & $-1.010$&$-0.967$ & $-0.980$ & $-0.986$ &  $-0.99\pm0.02$ \\
$r_0$ (fm) & 0.07&$-0.03$ & $-0.04$ & $-0.06$  & $-0.02\pm0.05$ \\
\hline
\end{tabular}
\end{center}
\end{table}

We also have to face uncertainties tied to the meson masses employed in our analysis.  Unlike in  \cite{hanhart_orginos}, the lattice spectrum used here is calculated with a pion mass of $m_{\pi}=156$~MeV, already very close to the physical value of 140~MeV. Moreover, since in the present case, only the kaon and $D,~D^*$ masses appear in the propagators and the potential is fitted to the lattice energy levels, there is no explicit dependence on $m_{\pi}$ in the analysis. We also assume that something similar occurs for the lattice energy levels and the changes between using 156~MeV or 140~MeV would be insignificant. This is actually the case for the chiral extrapolation of the $\bar K D$ and $K D$ scatttering lengths in \cite{hanhart_orginos}. However, the $D$ and $D^*$ masses of the lattice simulation are smaller than the physical ones, which is related to the Fermilab method employed (see $M_1$ in Table \ref{tableM}). This is the reason why we did not quote absolute values of the energies obtained, but the binding energies with respect to the thresholds. We can attempt to do an extrapolation of the results to physical masses. For this purpose we assume that the potential obtained can also be considered in absolute terms. Then we use this potential with the realistic masses in the loop function $G$ and obtain the results shown in Tables~\ref{extrap_1} and \ref{extrap_2}. 

\begin{table}[ht]
\caption{Extrapolation of the bound state properties to the physical mass of the $D$ meson, using $q_{\rm max}=1275$ MeV.}\label{extrap_1} 
\begin{center}
\begin{tabular}{l|c|c}
\hline
 $M_1$ (MeV) &  1631 & 1867 \\
 &  Ref.~\cite{sasa} & Physical \\
\hline
$B $(MeV) & 35.5 & 31.9 \\
$\mid g \mid$ (GeV) & 10.4 & 11.3 \\
$P$ (\%)  & 87.2 &  88.3 \\
$a_0$ (fm) & $-1.25$ & $-1.33$ \\
$r_0$ (fm) & $0.19$ & 0.14 \\
\hline
\end{tabular}
\end{center}
\end{table}

\begin{table}[ht]
\caption{Extrapolation of the bound state properties to the physical mass of the $D^*$ meson, using $q_{\rm max}=1275$ MeV.}\label{extrap_2} 
\begin{center}
\begin{tabular}{l|c|c}
\hline
 $M_1$ (MeV) &  1788 & 2008\\
 &  Ref.~\cite{sasa} & Physical \\
\hline
$B $(MeV) & 44 & 96 \\
$\mid g \mid$ (GeV) & 10.3 & 14.2 \\
$P$ (\%)  & 66.1 &  60.6 \\
$a_0$ (fm) & $-0.99$ & $-0.72$ \\
$r_0$ (fm) & $-0.060$ & $-0.002$ \\
\hline
\end{tabular}
\end{center}
\end{table}

 A third source of systematic uncertainties comes from the use of one type or another of the potentials, Eqs.~(\ref{pot}) or (\ref{eqcdd}), that we have already discussed in Sects.~\ref{aux} and \ref{sec:cdd}, respectively.  
Comparing the values given in Eqs.~(\ref{binding})-(\ref{P1ch}) with those of Eqs.~(\ref{binding2})-(\ref{P1chtwo}), we find that the systematic errors associated to the use of different potentials are:
\begin{eqnarray*}
&&\delta B(KD)=8.5~{\rm MeV} \ , \\
&&\delta B(KD^*)=7.5~{\rm MeV} \ , \\
&&\delta P(KD)=4.5~\% \ , \\
&&\delta P(KD^*)=4.0~\% \ , \\
&&\delta a(KD)=0.1~{\rm fm} \ , \\
&&\delta a(KD^*)=0.2~{\rm fm} \ , \\
&&\delta r_0(KD)=0.1~{\rm fm} \ , \\
&&\delta r_0(KD^*)=0.1~{\rm fm} \ .
\end{eqnarray*}

Altogether, summing these systematic errors in quadrature to those of Tables~\ref{qmax_1}-\ref{extrap_2}, we finally obtain the results: 
\begin{eqnarray*}
&&B(KD)=38 \pm 18 \pm 9~{\rm MeV} \ , \\
&&B(KD^*)=44 \pm 22 \pm 26~{\rm MeV} \ , \\
&&P(KD)=(72 \pm 13 \pm 5)~\%\ , \\
&&P(KD^*)=(57 \pm 21 \pm 6)~\%\ , \\
&&a(KD)=-1.3 \pm 0.5 \pm 0.1~{\rm fm} \ , \\
&&a(KD^*)=-1.1 \pm 0.5 \pm 0.2~{\rm fm} \ , \\
&&r_0(KD)=-0.1 \pm 0.3 \pm 0.1~{\rm fm} \ , \\
&&r_0(KD^*)=-0.2 \pm 0.3 \pm 0.1~{\rm fm} \ ,
\end{eqnarray*}
where the first error is statistical and the second systematic, which should also add in quadrature.

\section{Conclusions}
In this work we have done a reanalysis of the lattice spectra obtained in \cite{sasa,sasa1}  for {s-wave scattering channels $KD$ and $KD^*$, where bound states were   identified with the $D_{s0}^*(2317)$ and $D_{s1}^*(2460)$ states}. The analysis of \cite{sasa,sasa1} derived the scattering length and the effective range from two of the energy levels. The information of the third level was not used. Here we have done a reanalysis of the lattice spectra that takes into account the information of the three levels. The essence of the new method was the use of an auxiliary potential which was allowed to be energy dependent in the case of considering only one channel. This is demanded to take into account the fact that the single channels will most probably not  saturate the states. We found a bound state for both $KD$ and $KD^*$ scattering, which we associated to the $D_{s0}^*(2317)$ and $D_{s1}^*(2460)$ states.

In order to find out the most likely missing channels we were guided by the results of the chiral unitary approach which determines the $\eta D_s$, and $\eta D_s^*$ channels as the additional most important ones to saturate the wave function. However, the limited information from the lattice spectra drove us to use an energy independent potential with the consequence that the two channels chosen would saturate the wave function. With this restriction we found no solution, indicating that the lattice spectra does not contain information on the $\eta D_s$, and $\eta D_s^*$ channels. This seems to be the case since the levels found in \cite{sasa} are largely tied to the interpolators used, and no interpolators accounting for $\eta D_s$ and $\eta D_s^*$ states were included.

 We analyzed the lattice spectra considering only one channel and two energy dependent potentials.  One potential is taken linear in $s$ and another one contains a CDD pole accounting for possible genuine ${\bar c}s$ components. The results with both methods were compatible within errors. We also studied systematic uncertainties from other sources, which were found, in all cases but one, reasonably smaller than the statistical errors. Our analysis {confirmed}  the existence of bound states for the $KD$ and $KD^*$ channels with a binding of the order of 40 MeV, which we associated to the $D_{s0}^*(2317)$ and $D_{s1}^*(2460)$ states. We could also determine the scattering length and effective range for $KD$ and $KD^*$ scattering, improving on the previous results of \cite{sasa} based on the information of the lowest two levels only and relying upon the effective range formula. Finally, we could determine within errors that the states found are mostly of meson-meson nature and, using a sum rule which reformulates the test of compositeness condition of Weinberg, we established the probability to find $KD$ and $KD^*$ in those states in an amount of about 
$(72 \pm 13 \pm 5)~\%$ and $(57 \pm 21 \pm 6)~\%$, respectively. 
We discussed that, in order to be more precise on these numbers and obtain information on the channels that fill the rest of the probability, one must improve on the precision of the energy spectra and must include further interpolators that allow one to include the $\eta D_s$ and $\eta D_s^*$ channels in the analysis. 

  The exercise done shows the power of the method and the valuable information contained in the lattice spectra. The errors obtained here can be improved by having extra accuracy in the lattice spectra, additional levels, or more easy perhaps, spectra calculated for other lattice sizes. In any case, it has become clear that the information provided by the lattice spectra, and the flexibility to use different box sizes to obtain a rich spectrum of energies, is most useful when it comes to determine the energy dependence of the auxiliary potentials, which is essential to determine probabilities of meson meson components (or hadron hadron components in general)  via the generalized sum rule.

\section*{Acknowledgments}
 
S.P. is grateful to  C.B. Lang, L. Leskovec, D. Mohler and R. Woloshyn for the pleasant collaboration on the simulation that is analyzed in the present work. We would all like to thank them, and also M. D\"oring, for the subsequent useful discussions.  A. M. T  would like to thank the Brazilian funding agency FAPESP for the financial support. This work is partly supported by the Spanish Ministerio de Economia y Competitividad and European FEDER funds under contract numbers
FIS2011-28853-C02-01 and FIS2011-28853-C02-02,  by the Generalitat
Valenciana in the program Prometeo II, 2014/068, and by Grant 2014SGR-401from the Generalitat de Catalunya.
We acknowledge the
support of the European Community-Research Infrastructure
Integrating Activity Study of Strongly Interacting Matter (acronym
HadronPhysics3, Grant Agreement n. 283286) under the Seventh
Framework Programme of EU.

\bibliographystyle{plain}

\end{document}